\documentclass{article}

\usepackage{graphicx}
\newcommand{\ket}[1]{{|#1\rangle}} \newcommand{\bra}[1]{{\langle#1|}}

\author{Michele Mosca \quad and \quad Christof Zalka \\ \\
Department of Combinatorics and Optimization \\
University of Waterloo, Waterloo, Ontario \\
Canada N2L 3G1 \\
\small e-mail: {\tt
mmosca@iqc.ca} \quad and \quad {\tt zalka@iqc.ca}}

\title{Exact quantum Fourier transforms and discrete logarithm algorithms}

\begin{document}

\maketitle

\begin{abstract}
We show how the quantum fast Fourier transform (QFFT) can be made
exact for arbitrary orders (first for large primes). For most quantum
algorithms only the quantum Fourier transform of order $2^n$ is
needed, and this can be done exactly. Kitaev \cite{kitaev} showed how
to approximate the Fourier transform for any order. Here we show how
his construction can be made exact by using the technique known as
``amplitude amplification''. Although unlikely to be of any practical
use, this construction e.g. allows to make Shor's discrete logarithm
quantum algorithm exact. Thus we have the first example of an exact
non black box fast quantum algorithm, thereby giving more evidence
that ``quantum'' need not be probabilistic.

We also show that in a certain sense the family of circuits for the
exact QFFT is uniform. Namely the parameters of the gates can be
calculated efficiently.
\end{abstract}

\section{Introduction}

The ``quantum fast Fourier transformation'' (QFFT) plays an important
role in quantum algorithms. It is a unitary transformation that
applies the discrete Fourier transform to the amplitudes of a quantum
register. The standard version has order $2^n$ and is applied to a
quantum register consisting of $n$ qubits. It was found by Coppersmith
\cite{coppersmith} (see also Shor \cite{shor}). The construction is
essentially identical to the standard classical fast Fourier transform
(FFT). Like the FFT it generalises to orders which are a power of a
small prime and more generally to smooth numbers, thus integers who
have only small prime factors (see Cleve \cite{cleve2}). These
constructions implement the desired unitary transformation exactly.

In contrast, so far no exact (and efficient) constructions for
arbitrary orders have been known. For his ``Abelian stabiliser
problem'' Kitaev \cite{kitaev} gave an approximate implementation
based on ``eigenvalue estimation''. Here we show how this eigenvalue
estimation step can be made exact using ``amplitude
amplification''. Amplitude amplification \cite{ampamp} is a slight
generalisation of Grover's algorithm, allowing to apply the square
root speed up to any heuristic algorithm. Brassard and H{\o}yer
\cite{brassard} used a variant of it to make Simon's algorithm exact.

Finally we point out that an exact quantum Fourier transform for
large prime orders can be used to make Shor's discrete logarithm
algorithm exact.

\section{The exact QFFT$_p$ for large prime $p$}

The quantum Fourier transform of order (or ``modulus'') $N$ acts on
``computational'' basis states $\ket{x}$ as follows:
$$
\mbox{QFFT}_N : \qquad \ket{x} \quad \to \quad \ket{\Psi_x} =
\frac{1}{N} \sum_{y=0}^{N-1} e^{2 \pi i \frac{x y}{N}} \ket{y}.
$$
For arbitrary, in particular non-smooth N, Kitaev \cite{kitaev}
proposes to do this in two steps (second part of section 5 in
\cite{kitaev}, see also the review by Jozsa \cite{jozsa}):
$$
\ket{x} \quad \to \quad \ket{x,\Psi_x} \quad \to \quad \ket{\Psi_x}
$$
where, as usual, registers that ``appear out of nowhere'' are
understood to have been initialised in the standard state
$\ket{0}$. Similarly in the second step, one of the registers is
reset to this state and can thus again be left away.

The first step constructs the Fourier state $\ket{\Psi_x}$ for a given
$x$. This can be done exactly by first obtaining the ``uniform
amplitude'' superposition $\ket{\Psi_0}$ of the first $p$ basis states
of a register and then ``rephasing'' it:
\begin{equation} \label{prep}
\ket{x,0} \quad \to \quad \ket{x,\Psi_0} \quad \to \quad
\ket{x,\Psi_x} .
\end{equation}
As pointed out by Kitaev, $\ket{\Psi_0}$ can be obtained from
$\ket{0}$ by a sequence of SO(2) rotations applied to each qubit
in order from high to low significance, whereby the rotation angle
has to be controlled by the previously touched qubits. The
rephasing then simply consists of a rephasing on each qubit,
proportional to $x$ and the place value of the qubit.

The second step of Kitaev's construction is the reverse of
$$
\ket{\Psi_x,0} \quad \to \quad \ket{\Psi_x,x}.
$$
This is done through a technique known as ``eigenvalue
estimation'' (see also the article by Cleve et al. \cite{cleve}),
which details how to find the eigenvalue of an unknown eigenstate
of some unitary $U$. We will describe this in more detail later.
Here we only need to note that although this operation is not
exact, it leaves the eigenstate $\ket{\Psi_x}$ unchanged. Thus it
does:
\begin{equation} \label{pest}
\ket{\Psi_x,0} \quad \to \quad \ket{\Psi_x} \sum_{x'} c_{x,x'}
\ket{x',g_{x,x'}}
\end{equation}
where on the right hand side the superposition should be dominated by
the term with $x'=x$, such that a measurement would yield $x$ with
good probability. We also included some (unwanted) ``garbage''
$g_{x,x'}$ which may be produced along with the eigenvalue.

\subsection{Using amplitude amplification}

We now use ``amplitude amplification'' \cite{ampamp} to eliminate all
but the desired term $\ket{x,g_{x,x}}$.  We give here a quick review of
this generalisation of Grover's algorithm. We are given a unitary
operator $A$ which, when applied to the initial state $\ket{0}$, gives
an output state which has some component in a ``good'' subspace. Thus
the probability $|P_{good}~ A \ket{0}|^2$ is not too small, where
$P_{good}$ is the projector onto the good subspace. The amplitude
of the good component can be increased through the following procedure
$$
\left[ ~A ({\bf 1}+(e^{i \phi} -1) \ket{0}\bra{0}) A^{-1}~
({\bf 1}+(e^{i \varphi}-1) P_{good})~ \right]^T  A \ket{0}
$$
where the sequence of operations in the brackets is repeated $T$ times,
depending on the ``success probability'' of the ``algorithm'' $A$
alone. As in Grover's algorithm, the fastest increase is achieved when
both phases are chosen $\phi=\varphi=\pi$. The algorithm can be
analysed by noting that the state always remains in a subspace
spanned by the state we are seeking $P_{good}~ A \ket{0}$ and by
$(1-P_{good})~ A \ket{0}$. Usually an integer number of iterations
will not lead exactly to the desired state and so we need to chose
different (non-optimal) phases, either in all steps or only in the
last one or two. In our case we will leave the phases at their
standard settings, but will modify $A$ so that its success probability
is reduced to 1/4 where a single iteration leads exactly to the
desired state.

The operator $A$ will be given by eq. \ref{pest}, where the state
$\ket{\Psi_x}$ will have to be added as a ``spectator'' that is not
changed.

\subsubsection{``Recognising'' the correct solution}

Apart from the ``heuristic'' algorithm $A$, amplitude amplification
requires a way to ``recognise'' the good states. More precisely, we
need a way to apply the phase $e^{i \varphi}$ to the good subspace and
leave its orthogonal complement unchanged. So how can we check whether
a number $x'$ is the right eigenvalue of $\ket{\Psi_x}$, thus whether
$x'=x$? This can be done because the eigenstate $\ket{\Psi_x}$ is
still available exactly. Thus given a state of the form $\ket{\Psi_x}
\sum_{x'} c_{x,x'} \ket{x',g_{x,x'}}$, we can check the second
register against the first one. To do this we apply the reverse of the
steps in eq. \ref{prep} to these two registers, thus:
$$
\ket{x',\Psi_x} \quad \to \quad \ket{x',\Psi_{x-x'}} \quad \to \quad
\ket{x',\theta_{x-x'}}
$$
where in the second step we only act on the second register. The state
$\ket{\Psi_0}$ is mapped back to $\ket{0}$, while for $x' \not =x$ we
get some state $\ket{\theta_{x-x'}}$ orthogonal to $\ket{0}$.  We can
now apply the phase $e^{i \varphi}$ to the $\ket{0}$ state and undo
the previous operations.

\subsection{``Uniformising'' the success probability}

One obstacle to using amplitude amplification to make algorithms
exact is that the success probability of the ``heuristic''
algorithm $A$ must be known. But this probability may depend on
the (unknown) instance of the problem. In our case the success
probability of eigenvalue estimation on $\ket{\Psi_x}$ indeed does
depend on $x$. We can fix this problem by modifying $A$ such that
the new success probability will become instance independent and
equal to the average over all instances for the original $A$. To
do this uniformisation we pick an integer $r$ uniformly at random
from $\{0,1, \dots p-1\}$ and replace $\ket{\Psi_x}$ with
$\ket{\Psi_{x+r}}$, which is just a rephasing. We keep a record of
$r$ and subtract it again from the result of eigenvalue
estimation. To do this with a unitary $A$ we will need an
additional register for $r$, but this is no problem, as we have
already included the possibility that eigenvalue estimation (eq.
\ref{pest}) also generates some unwanted garbage $g_{x,x'}$.

So now exact amplitude amplification will allow us to do
$$
\ket{\Psi_x,0} \quad \to \quad \ket{\Psi_x} \ket{x,g_{x,x}}.
$$
To get rid of the ``garbage'' we can do the usual trick of copying the
wanted result $x$ into an additional ``save'' register and then
undoing the previous steps. In total this will lead to six
applications of $A$ for an exact QFFT.

In summary, the construction of an exact QFFT relies on making
eigenvalue estimation (on Fourier states $\ket{\Psi_x}$) exact. The
essential observations are that eigenvalue estimation leaves the
eigenstate  $\ket{\Psi_x}$ exactly unchanged and so it can be used for
the checking stage of amplitude amplification. Furthermore we used
that the success probability of estimating $x$ from $\ket{\Psi_x}$ can
rather easily be ``uniformised'' across all $x=0 \dots p-1$.

\section{An exact discrete logarithm algorithm}

An exact algorithm for the QFFT leads in a straightforward manner to
an exact algorithm for the discrete logarithm algorithm of the same
order. This was also observed for finite fields of prime order by
Brassard and H{\o}yer \cite{brassard} (Theorem 12). For smooth orders
(only small prime factors) the problem can easily be solved
classically. Here we give a quick review for the case when the order
is a large prime (see also \cite{proos}, section 2.2.3).

In a discrete logarithm problem we are given an element $\alpha$ which
generates a cyclic group of some finite order, here a prime.  Thus
$\alpha^p = e$. Then another element $\beta$ of the group is given and
we want to know which power of $\alpha$ it is; that is, the integer
$a$ for which $\beta=\alpha^a$. This is also written as $a=\log_\alpha
\beta$. In the quantum solution (see Shor \cite{shor}), we prepare two
registers, each in a uniform amplitude superposition of $p$ basis
states:
$$
\frac{1}{p} \sum_{x=0}^{p-1} \sum_{y=0}^{p-1} \ket{x,y}.
$$
Then we compute the function $\alpha^x \beta^y$ in an additional
register and measure it. This will leave the two registers in a
superposition of the form $\sum_y \ket{x_0-a \cdot y,y}$ where all
arithmetic operations are understood to be modulo $p$, $x_0$ is random
and $y$ runs over $0\dots p-1$. By Fourier transforming each register
with a QFFT$_p$ we get a similar state but without the offset $x_0$,
namely an equally weighted superposition of all states of the form
$\ket{x,a \cdot x}$ with $x=0\dots p-1$. A measurement will now allow
to compute $a$ in all cases except when $x=0$. Thus we have the known
and instance independent success probability of $1-1/p$, which allows
to easily make the algorithm exact by using (exact) amplitude
amplification.

\subsection{Alternatively: directly uniformising dlog}

Actually one can directly make the success probability of the dlog
algorithm instance independent. Thus one uses the usual algorithm with
a QFFT$_{2^n}$, but replaces $\beta$ with $\beta \cdot \alpha^r$ where
$r$ is again chosen uniformly at random from $0\dots p-1$. We have
noted this approach a while ago, but were not able to show that the
(now averaged) success probability can be computed efficiently, thus
it is not clear whether the circuit for a given $p$ can be computed
efficiently.

\section{Eigenvalue estimation}

In our case we want to estimate the eigenvalue of $\ket{\Psi_x}$
under the (unitary) cyclic shift operator $U$ which acts on
computational basis states as: $\ket{x} \to \ket{(x+1)
\bmod p}$. For eigenvalue estimation we need to do large
powers of $U$, which in this case is easy. Namely we first prepare
an auxiliary $n$-qubit register in a uniform amplitude
superposition of all its $N=2^n$ basis states. (We will choose $N$
to be larger than $p$, see below.) Then we do:
$$
\frac{1}{\sqrt{N}} \sum_{y=0}^{N-1} \ket{y}~ \ket{\Psi_x} \quad
\to \quad \frac{1}{\sqrt{N}} \sum_{y=0}^{N-1} \ket{y}~ U^y
\ket{\Psi_x} \quad = \quad \frac{1}{\sqrt{N}} \sum_{y=0}^{N-1}
e^{-2 \pi i \frac{x y}{p}} \ket{y}~ \ket{\Psi_x}.
$$
where we used that the eigenvalue of $\ket{\Psi_x}$ under $U$ is
$e^{-2 \pi i x/p}$. Note that the operation we have to do is simply a
modular addition on computational basis states, thus $\ket{a,b} \to
\ket{a,(a+b) \bmod p}$. After a Fourier transform of size $2^n$
on the auxiliary register, the probability of measuring $y$ would be
given by:
$$
p_y = f^2 (y-x N/p) \qquad \mbox{where} \quad
f(z) = \frac{\sin(\pi z)}{N \sin(\pi z /N)} .
$$
\begin{figure}[tbp]
\begin{center}
\includegraphics[scale = 1]{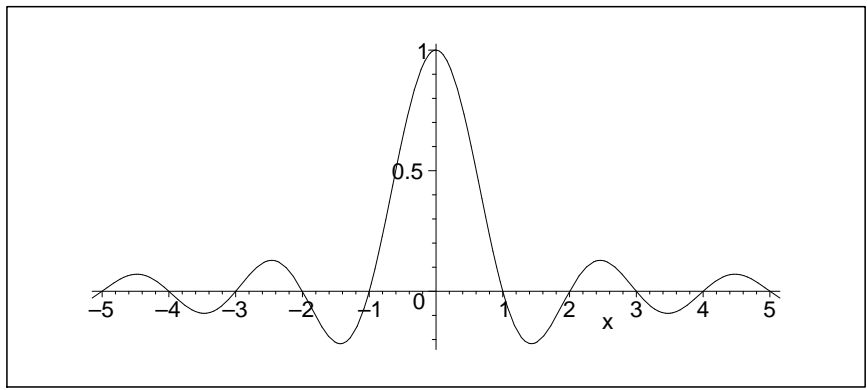}
\caption{\label{sinc} The function $\frac{\sin(\pi x)}{\pi x}$.}
\end{center}
\end{figure}
We illustrate the function $f(z)$ with $N \rightarrow \infty$ in
figure \ref{sinc}. It is peaked around $z=0$ so that after
measuring some $y$ we would guess for the number we want to find
$x \approx y\cdot p/N$. The choice with the highest probability of
obtaining the correct $x$ would be to simply round $y \cdot p/N$
to the closest integer. Partially to simplify notation, here we
round up to the next integer, thus our guess is $x'=\lceil y \cdot
p/N \rceil$. (For us the loss of some success probability does not
matter, at least not as long as it is at least 1/4.)

Because we have $N > p$, it is clear that if we measure $y=
\lfloor x \cdot N/p \rfloor$ we will calculate the correct $x$.
For a given $x$, smaller $y$ may also lead to the correct $x$, but
here we would like to eliminate this contribution to the success
probability, as it will lead to a simpler expression. Given a $y$
it is possible to eliminate these cases by also checking $y >
\lceil y \cdot p/N \rceil \cdot N/p -1$ and ``throwing away''
$y$'s which do not satisfy this. (Note that in order to obtain an
algorithm $A$ with a certain success probability, we can think as
if this were a non-reversible algorithm including measurements and
classical computations. Such an algorithm can then easily be
turned into a unitary $A$ which, besides the intended answer, also
produces some ``garbage''.) So now the success probability $p_x$
for correctly getting $x$ from $\ket{\Psi_x}$ is:
$$
p_x = f^2 (\lfloor x N/p \rfloor - x N/p) = f^2 ( \frac{x N \bmod p}{p}).
$$
To get the instance independent success probability of the uniformised
algorithm, we average this over all $x=0 \dots p-1$:
$$
\bar p = \frac{1}{p} \sum_{x=0}^{p-1} f^2 ( \frac{x N \bmod
p}{p}) = \frac{1}{p} \sum_{k=0}^{p-1} f^2 (k/p)
$$
where we have used that $N$ and $p$ are coprime and so for each
$x$ there is exactly one $k$.

\subsection{Efficiently calculating the success probability}

For large $p$ this sum of course is well approximated by the
corresponding integral, which (for large $N$) is approximately
0.4514. Here we show that for each $p$ and $N$, the success
probability can be approximated efficiently in the sense that the
computation time is polynomial in the number of (e.g. decimal) digits
we want to compute. The following method achieves this in a simple
way, although it is probably not the best one could do. Note that
$f^2(z)$ can be expanded in a (fast converging) power series in $x$
and $1/N$. (To compute $\bar p$ to $d$ digits we will only use
polynomially many terms in $d$.) Now each power $z^m$ of $z$ can be
summed separately, giving:
$$
\frac{1}{p}~ \sum_{k=0}^{p-1} (k/p)^m ~=~ \frac{1}{p^{m+1}}~ S_m (p) ~=~
\frac{1}{p^{m+1}}~ \sum_{i=0}^{m+1} A_{m,i}~ p^i 
$$
where for each power $m$ the coefficients $A_{m,i}$ can be calculated
(in various ways) in time polynomial in $m$. (A straightforward way
is to simply solve the equations resulting from $S_m(p+1)-S_m(p)=p^m$
for the $A_{m,i}$. E.g. for $m=1$ we get the familiar formula
$\sum_{k=0}^{p-1} k = p(p-1)/2$.)

\subsection{Adjusting the success probability}

Once we have calculated the success probability $\overline{p}$ (to
arbitrary precision) for a given $p$, we can use this to modify the
algorithm $A$ so that it will succeed exactly with probability 1/4, so
that just one iteration of amplitude amplification leads to an exact
algorithm. One way to do this is to add a qubit prepared in state
$\cos(\alpha) \ket{0}+\sin(\alpha) \ket{1}$ with $\overline{p}
\sin^2(\alpha) = \frac{1}{4}$ and additionally require for success
that this qubit be in state $\ket{1}$. The preparation of this qubit
will now require the one ``strange'' gate in our algorithm, although
its rotation angle $\alpha$ can be computed efficiently in the above
sense.

\section{Further remarks and observations}

\subsection{Generalisation to arbitrary orders}

The construction of the exact QFFT$_q$ easily generalises to arbitrary
orders $q$. Above we only needed the primality of the order for
(efficiently) computing the success probability. And there we only
needed that $N=2^n$ and $q$ should be coprime. Things can easily be
adjusted for the case when $q$ is even. Either we can modify (a bit)
the calculation of the success probability, or we can consider the
QFFT$_q$ as a tensor product of a QFFT with odd order and a standard
one with order a power of 2. Similarly, of course, we can generalise
to QFFT's over finite Abelian groups, not just cyclic ones.

Also the discrete logarithm algorithm can be generalised to arbitrary
orders $q$. Given the exact QFFT$_q$, the algorithm will be successful
whenever the first number in the measured pair $(x,x a \bmod q)$
is coprime to $q$. So the success probability is $\phi(q)/q$ where
$\phi(q)$ is the Euler totient function. If we know the factorisation
of $q$, this is easily calculated and so amplitude amplification can
be used to make the algorithm exact.

\subsubsection{Factorisation of the order of the dlog not known}

In the following we give a more involved solution for the case when
the factorisation of the order $q$ is not known. It consists of
$O(\log q)$ runs of (variants of) the dlog quantum circuit. What is
important is, that these variants still only use the special gates
calculated (efficiently) at the beginning from $q$.

In the first run it is enough, as before, to use amplitude
amplification only to get rid of the case $x=0$. We now measure a pair
$(x,x a \bmod q)$. If $x$ is coprime to $q$ we can directly calculate
$a$ and are done. If $\gcd(x,q)=d>1$, we still get some information
about $a$, namely $a'=a \bmod q/d$, and of course the factor $d$ of
$q$. Now we have $a=a'+a'' \cdot q/d$, where, in a standard way, $a''$
can be found by solving the dlog problem with $\tilde \alpha =
\alpha^{q/d}$ and $\tilde \beta = \beta \alpha^{-a'}$. This dlog
problem has smaller order, as $\tilde \alpha^d =e$, but we want to
reuse the original quantum circuit for order $q$. If in this original
circuit we simply replace $\alpha, \beta$ with $\tilde \alpha, \tilde
\beta$, we get (after the two QFFT's):
$$
\frac{1}{\sqrt{d}} \sum_{k=0}^{d-1} ~\ket{k \cdot q/d, ~a'' k \cdot
q/d~ \bmod q}
$$
(Note that this is essentially the same as $\sum_k \ket{k, a'' k \bmod
d}$.)  We want to avoid only the case $k=0$, but in order not to introduce
new ``special'' gates, we prefer to eliminate 3/4 of all states, such
that one step of standard amplitude amplification will lead to an
exact solution. We can e.g. only retain the last quarter of the values
$k=0\dots d-1$, although, if d is not divisible by 4, we will have to
``partially tag'' some states. (This can be done by appending a qubit
in state $c \ket{0}+s \ket{1}$ with $|s|^2=1/4,1/2$ or $3/4$.)

Now, like in the first step, we will either directly get $a''$, or
will gain partial information on $a''$, together with a factor of
$d$. This can be iterated (at most $O(\log q)$ times) till the order
of the dlog problem is small.

Note that in our construction we have taken care not to introduce new
``special'' gates during the computation. This means that really the
$O(\log q)$ quantum runs can be put together into one quantum circuit
whose gates can be computed from $q$ alone (without knowing its
factorisation).

\subsection{No exact factorisation algorithm}

Let us also note that it is not clear how to make Shor's integer
factorisation algorithm exact with the techniques used here. Thus
this is a challenge that remains. We note that Mosca \cite{mosca}
shows how to make factorisation exact in a slightly generalised
model of exact quantum computation.

\subsection{Review of other work on quantum Fourier transforms}

It is interesting to note that after Kitaev \cite{kitaev} a more
efficient and probably also more natural way to approximate the QFFT
for arbitrary orders has been given by Hallgren and Hales
\cite{hallgren}. In particular their construction uses fewer qubits,
but it seems not to lend itself to the techniques used here to make it
exact.

Also note the simplified ``semiclassical'' version of the standard
QFFT by Griffiths and Niu \cite{grif}. For practical implementations
of Shor's algorithms this would probably be the method of choice.

\subsubsection*{Acknowledgements}

The two authors have independently found the results described here,
although Ch.Z. acknowledges inspiration by related work of M.M. on
using amplitude amplification to make algorithms exact. Ch.Z. would
like to thank Professor David Jackson of our department for
discussions on summing the m$^{\tt th}$ powers of the first $n$
integers. He is supported by CSE (Communications Security
Establishment) and MITACS (Mathematics of Information Technology and
Complex Systems), both from Canada. M.M. thanks Richard Cleve, Lisa
Hales, and John Watrous for discussions at MSRI that prompted him to
think about solving the QFFT exactly in this context. M.M. is the
Canada Research Chair in Quantum Computation and is supported by
NSERC, MITACS, CFI, ORDCF, and PREA.

\end{document}